\documentclass[10pt]{article}

\usepackage{graphicx}              % to include figures
\usepackage{amsmath}               % great math stuff
\usepackage{amssymb, amsbsy, latexsym} % extra math symbols
\usepackage{amsfonts}              % for blackboard bold, etc
\usepackage[mathscr]{eucal}     % for \mathscr{...} font
\usepackage{amsthm}                % better theorem environments
\usepackage{multirow}               % allows an extra possibility in array environments
\usepackage{txfonts}

\DeclareMathSymbol{\C}{\mathbin}{AMSb}{"43}
\DeclareSymbolFont{AMSb}{U}{msb}{m}{n}

\makeatletter
\@addtoreset{equation}{section}
\makeatother

\newtheorem{Theorem}{Theorem}[section]
\newtheorem{Definition}{Definition}[section]

\def\be{\begin{eqnarray}}
\def\ee{\end{eqnarray}}

\newcommand{\ca}{\mathcal A}

\newcommand{\cd}{\mathcal D}

\newcommand{\ch}{\mathcal H}
\newcommand{\ci}{\mathcal I}

\newcommand{\cm}{\mathcal M}
\newcommand{\cn}{\mathcal N}

\newcommand{\cp}{\mathcal P}

\newcommand{\cs}{\mathcal S}

  \newcommand{\Fa}{\mathfrak{A}}

\newcommand{\fg}{\mathfrak{g}}

\newcommand{\g}{\gamma}

\newcommand{\eps}{\epsilon}

\newcommand{\Sig}{\Sigma}
\renewcommand{\l}{\lambda}
\renewcommand{\L }{\Lambda}
\renewcommand{\o}{\omega}
\renewcommand{\O}{\Omega}
\renewcommand{\t}{\tau}

\newcommand{\rmd}{\mathrm d}

\newcommand{\lt}{\left}
\newcommand{\rt}{\right}
\newcommand{\HN}{\hat{H}(N)}

\newcommand{\cyl}{\mathbf{Cyl}}
\newcommand{\lag}{\left\langle}
\newcommand{\rag}{\right\rangle}

\newcommand{\bbc}{\mathbb{C}}

\newcommand{\M}{\hat{\mathbf{M}}}

\begin{document}

\title{\sf A Path Integral for the Master Constraint of Loop Quantum Gravity}

\author{
{\sf Muxin Han}\thanks{{\sf
mhan@aei.mpg.de}}\\
\\
{\sf MPI f. Gravitationsphysik, Albert-Einstein-Institut,} \\
           {\sf Am M\"uhlenberg 1, 14476 Potsdam, Germany}\\
\\
{\sf Institut f. Theoretische Physik III, Universit\"{a}t Erlangen-N\"{u}rnberg} \\
{\sf Staudtstra{\ss}e 7, 91058 Erlangen, Germany}
}
%\date{{\tiny Preprint AEI-...}}
\date{}
\maketitle

\thispagestyle{empty}
\vspace{1.5cm}

\begin{abstract}

\vspace{0.3cm}

{\sf In the present paper, we start from the canonical theory of loop quantum gravity and the master constraint programme. The physical inner product is expressed by using the group averaging technique for a single self-adjoint master constraint operator. By the standard technique of skeletonization and the coherent state path-integral, we derive a path-integral formula from the group averaging for the master constraint operator. Our derivation in the present paper suggests there exists a direct link connecting the canonical Loop quantum gravity with a path-integral quantization or a spin-foam model of General Relativity. }

\end{abstract}

\newpage

\tableofcontents

\newpage

\section{Introduction}

Loop quantum gravity (LQG) is a mathematically rigorous quantization
of general relativity (GR) that preserves background independence
--- for reviews, see \cite{thiemann,book,rev}. It is inspired by the
formulation of GR as a canonical dynamical theory of connection
\cite{connection}. In this formulation, the canonical coordinate on
the phase space of GR is the su(2)-connections $A_a^i$ and the
densitized triads $E^a_i$. The total Hamiltonian of GR is a linear
combination of the Gauss constraint, the spatial diffeomorphism
constraint and the Hamiltonian constraint. Thus the dynamics of GR
are essentially the gauge transformations generated by the
constraints.

At the kinematical level of LQG, we smear the su(2)-connections
$A_a^i$ along paths $e$ to define the holonomies $A(e)$, and smear
the densitized triads $E_i^a$ on 2-surfaces to define the fluxes
$E(S)$. We consider an arbitrary finite piecewise analytic graph
$\g$ embedded in the spatial manifold $\Sig$, which is a collection
of edges (piecewise analytic paths) and their endpoints, called
vertices. We denote $E(\g)$ and $V(\g)$ the collections of the edges
and vertices of $\g$ respectively. A cylindrical function $f_\g$ is
a continuous function depending on the holonomies along the edges in
the graph $\g$. The collection of all the cylindrical functions
$\cyl$ with a certain sup-norm turns out to be an Abelian
$C^*$-algebra. The spectrum of this $C^*$-algebra is a compact
Hausdorff space \cite{conway}, which is denoted by $\overline{\ca}$
and called the quantum configuration space. The $C^*$-algebra of
cylindrical functions is the algebra of continuous functions on the
quantum configuration space $\overline{\ca}$. Moreover it turns out
that the fluxes $E(S)$ can be considered as the algebraic vector
fields on $\overline{\ca}$. The collections of the cylindrical
functions and the flux vector fields generates the so called,
holonomy-flux
*-algebra $\Fa$. The uniqueness theorem guarantees
that there exists a unique state (positive linear functional) on
this holonomy-flux *-algebra $\Fa$ such that it is invariant under
both $SU(2)$ gauge transformations and spatial diffeomorphism
transformations \cite{unique}. This invariant state gives a unique
representation Hilbert space $\ch_{Kin}$\footnote{In a way similar to that the states in Fock space are excitations from the unique (Poincar\'{a} invariant) vacuum state. } which is called the
kinematical Hilbert space of LQG. This kinematical Hilbert space is
isomorphic to the space of square-integrable function on
$\overline{\ca}$, $L^2(\overline{\ca},\rmd\mu_{AL})$, where the
measure $\mu_{AL}$ on $\overline{\ca}$ is called
Ashtekar-Lewandowski measure. This is the reason why we call
$\overline{\ca}$ the quantum configuration space. Moreover, the
space of cylindrical functions $\cyl$ is dense in the kinematical
Hilbert space $\ch_{Kin}$. A special class of cylindrical functions,
which is called the spin-network function $T_{s=(\g,j,m,n)}$, forms
an orthonormal basis in $\ch_{Kin}$. Since the spin-network function
$T_{s=(\g,j,m,n)}$ is labeled by a continuous parameter $\g$, the
kinematical Hilbert space $\ch_{Kin}$ is non-separable. Furthermore,
the Gauss constraint can be quantized and solved, which results in
the Hilbert space of $SU(2)$ gauge invariant states. The finite
spatial diffeomorphisms can be represented as unitary operators on
$\ch_{Kin}$. Thus by using the group averaging technique, we obtain
the Hilbert space of diffeomorphism states $\ch_{Diff}$. All above
is the beautiful kinematical framework of LQG, see \cite{thiemann}
for all the detailed constructions.

The dynamics of LQG is determined by the quantization of Hamiltonian constraint of GR. There are several different approaches to implement the Hamiltonian constraint at the quantum level. First of all, a quantum Hamiltonian constraint operator $\HN$ was first defined in \cite{QSD}. This operator is a densely defined closed operator on the kinematical Hilbert space $\ch_{Kin}$ and featured by its graph-changing operation on any cylindrical function $f_\g$. On the other hand, given a number of first-class constraints $C_I$, we can define a, so called, master constraint by
\be
\mathbf{M}:=\sum_{I,J}K^{IJ}C_IC_J\label{MASTER}
\ee
where $K$ is a positive definite matrix. Correspondingly, an operator $\hat{\mathbf{M}}$ is defined and usually, it is a positive self-adjoint operator. There are two versions of the master constraint operator in LQG. One is a positive self-adjoint operator defined on the Hilbert space $\ch_{Diff}$ of diffeomorphism invariant states, with the graph-changing operations \cite{master}. The other version is non-graph-changing, positive, self-adjoint operator, defined on the kinematical Hilbert space $\ch_{Kin}$. This non-graph changing master constraint operator is also adapted in the framework of algebraic quantum gravity (AQG) \cite{AQG}, which define the quantum theory of GR on a single cubic algebraic graph $\g$ with an infinite number of edges and vertices. Its kinematical Hilbert is a infinite tensor product extension \cite{ITP} of the LQG kinematical Hilbert space $\ch_{Kin}$. It is remarkable that the semiclassical limit of the non-graph changing master constraint is correctly reproduced by the complexifier coherent states \cite{hall,GCS,ce}. With the Hamiltonian constraint operator or the master constraint operator, a Quantum Einstein Equation is well-defined as the quantum constraint equation $\HN\Psi=0$ or $\M\Psi=0$\footnote{See section \ref{aqg} for the definition of $\M$.} in LQG, whose space of solutions with a chosen physical inner product is called physical Hilbert space $\ch_{Phys}$.

In order to rigorously define the physical Hilbert space from a self-adjoint master constraint operator, we follow the general procedure of direct integral decomposition (DID) \cite{DID}, which rigorously specify the physical inner product in general. This procedure is also tested in various physical models \cite{DID2} and gives correct results. Another approach to obtain the physical inner product is the group averaging technique for rigging inner product \cite{RAQ,link,link1}. In order to employ the group averaging technique for a given first-class constrained system, the constraint algebra formed by the first-class constraint should have a Lie algebraic structure so that the gauge transformations generated by the constraints form a group. In the case of GR, it is well-known that the constraint algebra formed by the diffeomorphism constraint and Hamiltonian constraint is not a Lie algebra, because of the presence of structure function. Thus the group averaging doesn't work directly for the diffeomorphism-Hamiltonian constraint algebra. However if we quantize the constraint system using a single master constraint, the constraint algebra is trivial thus is a trivial Lie algebra (since there is only one constraint). Therefore the group averaging can also be carried out for the master constraint, which has been already pointed out in \cite{master} long time ago. In addition, we have shown that the group averaging for the master constraint gives the consisent result as it is given by DID procedure under some technical assumptions \cite{link1}, which are fulfilled by all the physical models tested in \cite{DID}. So if the master constraint operator in LQG satisfies those assumptions, the group averaging technique is the correct way to compute the physical inner product in LQG.

The general consideration in the last paragraph is a preparation for linking canonical LQG with a path-integral formula. The path-integral formulation is of interest in the quantization of general relativity (GR), a theory where space-time covariance plays a key role. Currently the spin-foam model \cite{spinfoam2} can be thought of as a path-integral framework for loop quantum gravity, directly motivated from the ideas of path-integral adapted to reparametrization-invariant theories. The current spin foam approach is independent from the dynamical theory of canonical LQG
\cite{QSD} because the dynamics of canonical LQG is rather complicated, it uses
an apparently much simpler starting point:
Namely, in the Plebanski formulation \cite{plebanski}, GR can be
considered as a
constrained BF theory and treating the so called simplicity
constraints as a perturbation of BF theory, one can make use of the
powerful toolbox that come with topological QFT's \cite{bf}. It is
an unanswered question, however, and one of the most active research
topics momentarily\footnote{Here we are referring the spin-foam model for 4-dimensional gravity.}, how canonical LQG and spin foams fit together.
It is one the aims of this paper to make a contribution towards
answering this question.

There are some early attempts in looking for a link between canonical LQG and a path-integral formulation of LQG. In the seminal paper \cite{RR}, the authors presented a heuristic method in solving the (Euclidean) Hamiltonian constraint operator in LQG, i.e. taking any diffeomorphism invariant state $\psi\in\ch_{Diff}$ and map to
\be
\delta(\hat{H}')\psi:=\prod_{x\in\Sig}\delta(\hat{H}'(x))\psi=\int\cd N\ e^{i\int_\Sig\rmd^3x\ N(x)\hat{H}'(x)}\psi\label{rr}
\ee
where $\hat{H}'$ is the dual operator on $\cyl^\star$. Eq.(\ref{rr}) have a shape similar as a group averaging. However such a method is formal because (1.) The exponential is not well-defined because $\hat{H}'(x)$ is not a self-adjoint operator; (2.) the product the operator $\hat{H}'(x)$ doesn't preserve the $\ch_{Diff}$. If we use $\hat{H}^\dagger(x)$ and operating on kinematical state $\psi\in\ch_{Kin}$, the product $\prod_{x\in\Sig}$ is not well-defined since for different $x\in\Sig$, $\hat{H}^\dagger(x)$ doesn't commute. However if we proceed formally, we obtain the physical inner product for given two diffeomorphism invariant spin-network states
\be
\lag T_{[s]},T_{[s']}\rag_{Phys}=\sum_{n=0}^\infty\frac{i^n}{n!}\int\cd N\ T_{[s']}\lt[\hat{H}^\dagger(N)^nT_s\rt]
\ee
The physical inner product so defined has a spin-foam-like structure since the Hamiltonian constraint operator $\hat{H}^\dagger(N)$ is graph-changing, i.e. add several arcs in the neighborhood of each vertex whose valence more than 3. But as we have seen, although this physical inner product suggests the spin-foam as the path-integral quantization of GR, it is unfortunately ill-defined, essentially by the fact that the constraint algebra formed by the Hamiltonian constraint is not a Lie-algebra.

The present paper applies the general considerations in \cite{link,link1} to the case of GR. Here we improve the situation in Eq.(\ref{rr}) by using a single self-adjoint master constraint $\M$. As it is shown in \cite{link1}, the group averaging using the master constraint operator is well-defined and correctly specifies the physical inner product and physical Hilbert space under some technical assumptions\footnote{More precisely, the group averaging using the master constraint operator correctly specifies the absolutely continuous sector of the physical Hilbert space under some technical assumptions.}. In this paper, we will show that this group averaging physical inner product can be computed practically (under some technical assumptions) by using the available semiclassical technique \cite{GCS}. It turns out that the result of the computation lead to a path-integral formula of LQG, which is a discrete analog of the naive Ansatz:
\be
\int\cd\mu\ e^{iS_{GR}}=\int\cd A_a^j\cd E^a_j\cd\L^j\cd N^a\cd N \exp\lt[-\frac{i}{\ell_p^2}\int\rmd t\rmd^3x\lt(E^a_j\partial_tA^j_a-\L^jG_j-N^aH_a-NH\rt)\rt]\label{naive}
\ee
up to a local measure factor when the matrix $K$ in Eq.(\ref{MASTER}) is phase space dependent.

The present paper is organized as the follows:

In section \ref{aqg}, we briefly review the framework of algebraic quantum gravity and the master constraint programme, the definition of group averaging technique and its consistency with direct integral decomposition. We also briefly review the semiclassical tools in AQG, which will be employed in derive the path-integral.

In section \ref{main}, we start from the definition of the group averaging physical inner product using the master constraint. By using the standard technique of skeletonization and coherent state path-integral, we derive a path-integral formula from the discrete setting. With some certain assumptions and approximations, we obtain a path-integral formula in analogy with Eq.(\ref{naive}) up to a local measure factor.

In section \ref{conc}, we summarize and conclude.

\section{Algebraic quantum gravity and master constraint programme}\label{aqg}

\subsection{The framework of algebraic quantum gravity}

Algebraic quantum gravity (AQG) \cite{AQG} is a modified approach of loop quantum gravity. But in contrast to loop quantum gravity, the quantum kinematics of AQG is determined by an abstract $*$-algebra generated by a countable set of elementary operators labeled by a single algebraic graph $\g$ with countably infinite number of edges\footnote{As it will explained later in this subsection, the Hilbert space on the single graph $\g$ contain rich enough information, more precisely the Hilbert space is non-separable, since the graph $\g$ is infinite.}, while in loop quantum gravity the elementary operators are labeled by a collection of embedded graphs with a finite number of edges. Thus one can expect that in AQG, we don't consider the information of the topological and differential structure of the manifold in all the quantization procedure except the semi-classical analysis. When we consider the semiclassical limit of AQG, we should specify an embedding $X:\ \g\to\Sig$, which makes the contact of the abstract operators with the physical fields on the spatial manifold $\Sig$.

The quantization in AQG bases on a single algebraic graph, which only contains the information of the number of vertices and their oriented valence.

\begin{Definition}
An oriented algebraic graph $\g$ is an
abstract graph specified by its adjacency matrix $\g$, which is
an $N\times N$ matrix. One of its entries $\g_{IJ}$ stand for
the number of edges that start at vertex $I$ and end at vertex $J$.
The valence of the vertex $I$ is given by
$v_I=\sum_J(\g_{IJ}+\g_{JI})$. We also use $V(\g)$ and
$E(\g)$ to denote the sets of vertices and edges respectively.
\end{Definition}

In our quantization procedure, we fix a specific cubic algebraic graph with a countably infinite number of edges $N=\aleph$ and the valence of each vertex $v_I=2\times \dim(\Sigma)$.

Given the algebraic graph $\g$, we define a quantum $*$-algebra by associating with each edge $e$ an element $A(e)$ of a compact, connected, semisimple Lie group $G$ and an element $E_j(e)$ take value in its Lie algebra $\mathfrak{g}$. These elements for all $e\in E(\g)$ are subject to the canonical commutation relations
\begin{eqnarray}
&&[\hat{A}(e),\hat{A}(e')]=0, \nonumber\\
&&[\hat{E}_j(e),\hat{A}(e')]=i\hbar Q^2\delta_{e,e'}\tau_j/2\hat{A}(e),\nonumber\\
&&[\hat{E}_j(e),\hat{E}(e')]=-i\hbar
Q^2\delta_{e,e'}f_{jkl}\hat{E}_l(e'),
\end{eqnarray}
and $*$-relations
\begin{eqnarray}
\hat{A}(e)^*=[\hat{A}(e)^{-1}]^T,\ \ \ \ \
\hat{E}_j(e)^*=\hat{E}_j(e),
\end{eqnarray}
where $Q$ stands for the coupling constant ($Q^2=\kappa$ in the case of GR), $\tau_j$ is the generators in the Lie algebra $\mathfrak{g}$ and $f_{jkl}$ is the structure constant of $\mathfrak{g}$. We denote the abstract quantum $*$-algebra generated by above elements and relations by $\mathfrak{A}$.

A natural representation of $\mathfrak{A}$ is the infinite tensor
product Hilbert space \cite{ITP}
\be
\ch_{Kin}:=\mathcal{H}^{\otimes}=\otimes_{e\in E(\g)}\mathcal{H}_e
\ee
where $\mathcal{H}_e=L^2(G,d\mu_H)$, whose element
is denoted by $\otimes_f\equiv\otimes_ef_e$. Two elements
$\otimes_f$ and $\otimes_{f'}$ in $\mathcal{H}^{\otimes}$ are said
to be strongly equivalent if $\sum_e|<f_e,f'_e>_{\mathcal{H}_e}-1|$
converges. We denote by $[f]$ the strongly equivalence class
containing $\otimes_f$. It turns out that two elements in
$\mathcal{H}^{\otimes}$ are orthogonal if they lie in different
strongly equivalence classes. Hence the infinite tensor Hilbert
space $\mathcal{H}^{\otimes}$ can be decomposed as a direct sum of
the Hilbert subspaces (sectors) $\mathcal{H}^\otimes_{[f]}$ which
are the closure of strongly equivalence classes $[f]$. Furthermore,
although each sector $\mathcal{H}^\otimes_{[f]}$ is separable and
has a natural Fock space structure, the whole Hilbert space
$\mathcal{H}^{\otimes}$ is non-separable since there are uncountably
infinite number of strongly equivalence classes in it. Our basic
elements in the quantum algebra are represented on
$\mathcal{H}^{\otimes}$ in an obvious way
\begin{eqnarray}
\hat{A}(e)\otimes_f&:=&[A(e)f_e]\otimes[\otimes_{e'\neq
e}f_{e'}],\nonumber\\
\hat{E}_j(e)\otimes_f&:=&[i\hbar Q^2X^e_jf_e]\otimes[\otimes_{e'\neq
e}f_{e'}].
\end{eqnarray}
where $X_j^e$ is the right-invariant vector field on the Lie group $G$.
As one might have expected, all these operators are densely defined
and $E_j(e)$ is essentially self-adjoint. Given a vertex $v\in
V(\g)$, the volume operator can be constructed by using the
operators we just defined
\begin{eqnarray}
\hat{V}_v:=\ell_p^3\sqrt{|\frac{1}{48}\sum_{e_1\cap e_2\cap
e_3=v}\epsilon_v(e_1, e_2,
e_3)\epsilon^{ijk}\hat{E}_i(e_1)\hat{E}_j(e_2)\hat{E}_k(e_3)|},
\end{eqnarray}
where the values of $\epsilon_v(e_1, e_2, e_3)$ should be assigned
once for all for each vertex $v\in V(\g)$. When we embed the algebraic graph $\g$ into
some manifold, the embedding should be consistent with the assigned
values of $\epsilon_v(e_1, e_2, e_3)$.

Then we discuss the quantum dynamics. The half densitized
constraints can be quantized to be composite operators as we list
below \cite{AQG}.
\begin{itemize}
\item Gauss constraint
\begin{eqnarray}
\hat{G}^{(1/2)}_j(v):=\hat{Q}_v^{(1/2)}\lt[\sum_{b(e)=v}\hat{E}_j(e)-\sum_{f(e)=v}O_{jk}\lt[\hat{A}(e)\rt]\hat{E}_k(e)\rt];
\end{eqnarray}

\item Spatial diffeomorphism constraint
\begin{eqnarray}
\hat{D}^{(1/2)}_j(v)&:=&\frac{1}{E(v)}\sum_{e_1\cap e_2\cap
e_3=v}\frac{\epsilon_v(e_1, e_2,
e_3)}{|L(v,e_1,e_2)|}\sum_{\beta\in
L(v,e_1,e_2)}\ \mathrm{tr}\lt(\tau_j\Bigg[\hat{A}(\beta)-\hat{A}(\beta)^{-1}\Bigg]\ \hat{A}(e_3)\lt[\hat{A}(e_3)^{-1},\sqrt{\hat{V}_v}\rt]\rt);
\end{eqnarray}

\item Euclidean Hamiltonian constraint (up to an overall factor)
\begin{eqnarray}
\hat{H}^{(r)}_E(v)&:=&\frac{1}{E(v)}\sum_{e_1\cap e_2\cap
e_3=v}\frac{\epsilon_v(e_1, e_2,
e_3)}{|L(v,e_1,e_2)|}\sum_{\beta\in
L(v,e_1,e_2)}\ \mathrm{tr}\lt(\Bigg[\hat{A}(\beta)-\hat{A}(\beta)^{-1}\Bigg]\ \hat{A}(e_3)\Bigg[\hat{A}(e_3)^{-1},\hat{V}_v^{(r)}\Bigg]\rt);
\end{eqnarray}

\item Lorentzian Hamiltonian constraint (up to an overall factor)
\begin{eqnarray}
\hat{T}(v)&:=&\frac{1}{E(v)}\sum_{e_1\cap e_2\cap
e_3=v}\epsilon_v(e_1, e_2,
e_3)\ \mathrm{tr}\lt(\hat{A}(e_1)\Bigg[\hat{A}(e_1)^{-1},\lt[\hat{H}_E^{(1)},\hat{V}\rt]\Bigg]\
\hat{A}(e_2)\Bigg[\hat{A}(e_2)^{-1},\lt[\hat{H}_E^{(1)},\hat{V}\rt]\Bigg]\ \hat{A}(e_3)\lt[\hat{A}(e_3)^{-1},\sqrt{\hat{V}_v}\rt]\rt),\nonumber\\
\hat{H}^{(1/2)}(v)&=&\hat{H}^{(1/2)}_E(v)+\hat{T}(v);
\end{eqnarray}

\end{itemize}
where the matrix $O_{jk}[g]$ is the adjoint representation of $g\in G$ on the Lie algebra $\fg$, $\hat{V}:=\sum_v \hat{V}_v$,
$\hat{H}_E^{(1)}:=\sum_v\hat{H}_E^{(1)}(v)$ and
\begin{eqnarray}
\hat{Q}_v^{(r)}&:=&\frac{1}{E(v)}\sum_{e_1\cap e_2\cap
e_3=v}\epsilon_v(e_1, e_2,
e_3)\ \mathrm{tr}\lt(\hat{A}(e_1)\lt[\hat{A}(e_1)^{-1},\hat{V}_v^{(r)}\rt]\ \hat{A}(e_2)\lt[\hat{A}(e_2)^{-1},\hat{V}_v^{(r)}\rt]\
\hat{A}(e_3)\lt[\hat{A}(e_3)^{-1},\hat{V}_v^{(r)}\rt]\rt).
\end{eqnarray}
$E(v)$ denotes the binomial coefficient which comes from
the averaging with respect to the triples of edges meeting at given
vertex $v$. $L(v,e_1,e_2)$ denotes the set of minimal loops starting at $v$
along $e_1$ and ending at $v$ along $e_2^{-1}$. And a loop $\beta\in
L(v,e_1,e_2)$ is said to be minimal provided that there is no other
loop within $\g$ satisfying the same restrictions with fewer
edges traversed. Note that since we only have a single cubic
algebraic graph, the diffeomorphism constraint can only be
implemented by defining the operators corresponding to
diffeomorphism generators because a finite diffeomorphism
transformation is not meaningful in our algebraic treatment unless
the algebraic graph is embedded in a manifold. As a result, the
(extended) master constraint operator  can be expressed as a quadratic
combination:
\begin{eqnarray}
\hat{\mathbf{M}}:=\sum_{v\in V(\g)}[\hat{G}^{(1/2)}_j(v)^\dagger
\hat{G}^{(1/2)}_j(v)+\hat{D}^{(1/2)}_j(v)^\dagger \hat{D}^{(1/2)}_j(v)+\hat{H}^{(1/2)}(v)^\dagger
\hat{H}^{(1/2)}(v)]\label{M}.
\end{eqnarray}
It is trivial to see that all the above operators are
non-graph-changing\footnote{The non-graph-changing operator means that the operator doesn't change the underline infinite cubic graph $\gamma$, i.e. it doesn't add any edge in addition to the edges in $E(\g)$.}
and embedding independent because we have only
worked on a single algebraic graph so far. It is obvious that the master constraint operator $\M$ is a positive and symmetric operator, thus has the FriedrichÕs self-joint extension \cite{reedsimon}. We take this self-adjoint extension of $\M$ and consider $\M$ as a self-adjoint operator.

Note that the master constraint operator $\M$ is not densely defined on the whole infinite tensor product Hilbert space $\ch_{Kin}$. In order to make $\M$ densely defined we have to restrict ourselves in a subspace such that $\M$ is densely defined on the subspace. On the other hand, it is remarkable that the master constraint operator $\M$ preserves all the strong equivalence class sectors, which are separable Hilbert spaces although the whole $\ch_{Kin}$ is non-separable. Because of this property, we can use the direct integral decomposition (DID) separately in each strong equivalence class sector to define the physical Hilbert space for each sector (see the next section). Then the total physical Hilbert space is a direct sum of the sectorial physical Hilbert spaces. Moreover, the strong equivalence class sector $\ch_{AL}$ where the vacuume state $\o=1$ lives is especially interesting. A generic states in the sector $\ch_{AL}$ only has excitations on a finite number of edges, which shows the similarity with the kinematical Hilbert space in LQG.

\subsection{Physical inner product and group averaging technique}

Given the self-adjoint master constraint operator $\M$ in Eq.(\ref{M}), we can formally define the quantum master constraint equation by
\be
\M\ \Psi=0\label{MEQ}
\ee
The space of solutions for this equation combined with a certain physical inner product is called the physical Hilbert space $\ch_{Phys}$. However, firstly the equation Eq.(\ref{MEQ}) is only formally written because zero is probably contained in the continuous spectrum of the master constraint operator, so that the solution state $\Psi$ may be not living in the kinematical Hilbert space anymore. Secondly it is not clear from Eq.(\ref{MEQ}) about which physical inner product should be chosen on the space of solutions, unless zero is a pure point spectrum and the space of solutions is a subspace of $\ch_{Kin}$. In order to rigorously define the physical Hilbert space $\ch_{Phys}$, we should in principle employ the direct integral decomposition (DID) \cite{DID} for the master constraint operator $\M$. Note that DID requires that the Hilbert space should be separable. Fortunately, since the master constraint operator $\M$ is self-adjoint and preserves all the strong equivalence class sector in the infinite tensor product Hilbert space, the physical Hilbert space $\ch_{Phys}$ is well-defined in principle for each strong equivalence class sector. Finally we should make a direct sum for all the resulting sectorial physics Hilbert spaces.

In \cite{DID}, the programme of direct integral decomposition (DID) is introduced in order to rigorously define the physical Hilbert space for a general constraint system. Such a programme proceeds as the follows:

\begin{enumerate}

\item Given a separable kinematical Hilbert space $\mathcal{H}_{Kin}$ and a self-adjoint master constraint operator $\mathbf{M}=K^{IJ}C^\dagger_IC_J$, first of all we have to split the kinematical Hilbert space into three mutually orthogonal sectors $\mathcal{H}_{Kin}=\mathcal{H}^{pp}\oplus\mathcal{H}^{ac}\oplus\mathcal{H}^{cs}$ with respect to the three different possible spectral types of the master constraint operator $\mathbf{M}$.

\item We make the direct integral decomposition of each $\mathcal{H}^*$, $*=pp,ac,cs$\footnote{It labels respectively pure point, absolutely continuous, or continuous singular spectrum of the self-adjoint operator. These names come from the properties of the spectral measures on the corresponding sectors. See \cite{DID} for detailed definition.} with respect to the spectrum of the master constraint operator $\M$ restricted in each sector, i.e.
\begin{eqnarray}
\mathcal{H}^*=\int^\oplus\rmd\mu^*(\l)\ch^\oplus_\l\nonumber
\end{eqnarray}

\item Finally we define the physical Hilbert space to be a direct sum of three fiber Hilbert spaces at $\lambda=0$ with respect to three kinds of spectral types, i.e. $\mathcal{H}_{Phys}=\mathcal{H}^{pp}_{\lambda=0}\oplus\mathcal{H}^{ac}_{\lambda=0}\oplus\mathcal{H}^{cs}_{\lambda=0}$
\end{enumerate}

Note that in step 2, we have assumed that all the ambiguities outlined in \cite{DID} have been solved by considering some physical criterions e.g. the physical Hilbert space should admit sufficiently many semiclassical states, and it should irreducibly represent the algebra of Dirac observables as an algebra of self-adjoint operators. With this assumption, the procedure of DID programme gives a precise definition of the physical Hilbert space for a general constraint system. And in many models simpler than GR, such a programme gives satisfactory physical Hilbert space \cite{DID2}.

However, if we want to practically obtain the physical Hilbert space of AQG and the detailed knowledges about the structure of this physical Hilbert space, then DID is not a suitable procedure. The reason is the following: The whole procedure of DID depends on a good knowledge of the spectral structure for the master constraint operator. While for the case of LQG or AQG with a complicated master constraint operator $\M$, the spectrum of $\M$ is largely obscure so that the DID programme is too hard to proceed practically even the first step. Therefore for our practical purpose in exploring the AQG physical Hilbert space, we have to employ a technique such that the final structure of physical Hilbert space $\mathcal{H}_{Phys}=\mathcal{H}^{pp}_{\lambda=0}\oplus\mathcal{H}^{ac}_{\lambda=0}\oplus\mathcal{H}^{cs}_{\lambda=0}$ (or some sectors in it) is written down at least formally without much of the knowledges for the spectrum of the master constraint operator. Fortunately we have a single constraint in the quantum theory, whose generated gauge transformations for a one-parameter group. Therefore we can employ an alternative, (modified) group averaging technique to obtain the physical inner product \cite{link1}:

\begin{Definition}\label{GA}
For each state $\psi$ in a dense subset $\mathcal{D}$ of $\mathcal{H}_{Kin}$, a linear functional $\eta_\Omega(\psi)$ in the algebraic dual of $\mathcal{D}$ can be defined such that $\forall \phi\in\mathcal{D}$
\begin{eqnarray}
\eta_{\Omega}(\psi)[\phi]:=\lim_{\epsilon\rightarrow0}\frac{\int_{\mathbb{R}}\mathrm{d}t\ \langle\psi|e^{it(\mathbf{M}-\epsilon)}|\phi\rangle_{Kin}}{\int_{\mathbb{R}}\mathrm{d}t\ \langle\Omega|e^{it(\mathbf{M}-\epsilon)}|\Omega\rangle_{Kin}}\nonumber
\end{eqnarray}
where $\Omega\in\mathcal{H}_{Kin}$ is called a reference vector. Therefore we can define a new inner product on the linear space of $\eta_\Omega(\psi)$ via $\langle\eta(\psi)|\eta(\phi)\rangle_\Omega:=\eta_{\Omega}(\psi)[\phi]$. The resultant Hilbert space is denoted by $\mathcal{H}_\Omega$
\end{Definition}

It turns out that the group averaging Hilbert space $\ch_\O$ is consistent with the physical Hilbert space defined by DID in a certain sense \cite{link1}:

\begin{Theorem} \label{GA1}
We suppose zero is not a limit point in $\sigma^{pp}(\mathbf{M})$ and $\sigma^{cs}(\mathbf{M})=\emptyset$ (which relates the argument that there is no state without physical interpretation), In addition, if we have any one of the following conditions
\begin{enumerate}
\item there exists $\delta>0$ such that each $\mu^{ac}_m$ ($\mathrm{d}\mu^{ac}_m=\mu^{ac}_m\mathrm{d}\lambda$) is continuous on the closed interval $[0,\delta]$.
\item there exists $\delta>0$ such that each $\rho^{ac}_m$ is continuous at $\lambda=0$ and is differentiable on the open interval $(0,\delta)$.
\item there exists $\delta>0$ such that $N^{ac}$ is constant on the neighborhood $[0,\delta)$.
\end{enumerate}
Then there exists a dense domain $\mathcal{D}$ in $\mathcal{H}_{Kin}$, such that for some certain choices of reference vector $\Omega$ the group averaging Hilbert space $\mathcal{H}_\Omega$ is unitarily equivalent to the absolute continuous sector of physical Hilbert space $\mathcal{H}^{ac}_{\lambda=0}$.
\end{Theorem}

See \cite{link1} for the proof of this theorem. Here we see that the reason of taking the limit $\epsilon\to0$ in Definition \ref{GA} is to make the desired connection between the group averaging Hilbert space $\mathcal{H}_\Omega$ and the absolute continuous sector $\mathcal{H}^{ac}_{\lambda=0}$ in physical Hilbert space. For the pure point sector $\mathcal{H}^{pp}_{\lambda=0}$, one should rather solve the eigen-equation Eq.(\ref{MEQ}) in the kinematical Hilbert space $\ch_{Kin}$. For the case of LQG or AQG, many eigenstates in $\ch_{Kin}$ have been found, which correspond to the degenerated geometry, e.g. the spin-networks with valence less than 4.

It is remarkable that all the physical models simpler than quantum gravity tested in \cite{DID2} satisfy all the assumptions in Theorem \ref{GA1}. It means that the group averaging technique in Definition \ref{GA} gives correct physical Hilbert space (the absolute continuous sector) for all those physical models. Therefore in applying the group averaging technique to the AQG master constraint operator $\M$, we will assume that the master constraint operator $\M$ defined in Eq.(\ref{M}) also satisfies all the assumptions in Theorem \ref{GA1}, in order to obtain the correct physical Hilbert space consistent with DID.

We will see in Section \ref{main} that we can proceed the practical computation for the group averaging inner product
\be
\int_{\mathbb{R}}\mathrm{d}t\ \langle\psi|e^{it(\mathbf{M}-\epsilon)}|\phi\rangle_{Kin}
\ee
by using a standard skeletonization procedure \cite{brown}. It turns out to be more convenient to use the coherent states in the skeletonization, which will be briefly reviewed in the next subsection. Finally, we will obtain a path-integral formula for the group averaging inner product.

\subsection{Coherent states in algebraic quantum gravity}

Before we derive the path-integral formula from the master constraint $\M$, we brief review the definition of coherent state in LQG, in order to use the method of coherent state path-integral in the next section. Since we are considering an algebraic graph $\g$ with cubic topology, it is shown that the LQG coherent state is qualified as a proper semiclassical states in the kinematical Hilbert space $\ch_{Kin}$. Its semiclassical expectation values for the geometrical operators and the master constraint operator reproduce GR correctly as $\hbar\to0$ \cite{GCS,AQG,ce}.

When we consider the semi-classical analysis of the above quantum framework, we should choose a embedding $X$ of the algebraic cubic graph $\g$ into spatial manifold $\Sig$, such that $\g$ is dual to the sub-2-complex $\cs$ in a cubic partition of $\Sig$. Thus for each $X(e)$ there is a face $S_e\in\cs$ which intersects $\g$ transversely only at an interior point $p_e$ of both $S_e$ and $X(e)$, for each $x\in S_e$, we choose a path $\rho_e(x)$ which starts in begin point of $X(e)$ and along $X(e)$ until $p_e$ and then runs within $S_e$ until $x$. We denote the path system for all $x\in S_e$ by $\cp_S$. Given the data $(\g,X,\cs,\cp)$, for a classical $G$-connection $A$ and a Lie($G$) valued vector density $E$ of weight one, as a pair of conjugate variables, we define the holonomy and the gauge covariant flux for each edge $e\in E(\g)$:
\be
h(e)&:=&\cp \exp \int_{X(e)}A\nonumber\\
P(e)&:=&\int_{S_e}\eps_{abc}\rmd x^a\wedge\rmd x^b\ h(\rho_e(x))\ E^c(x)\ h(\rho_e(x))^{-1}.\nonumber
\ee
And we define the dimensionless quantities $p_j=\frac{1}{a_e^2}P_j=-\frac{1}{2a_e^2}\mathrm{Tr}(\t_jP)$, $\t_j=-i\lt(\text{Pauli matrix}\rt)_j$ for $G$=SU(2), and $a_e$ is a parameter with dimension $[length]$. By using the fundamental Poisson bracket between $A^j_a$ and $E^a_j$, we can check that the holonomies $h(e)$ and fluxes $p_j(e)$ form the following Poisson algebra \cite{GCS}:
\be
\lt\{{h}(e),{h}(e')\rt\} &=&0\nonumber\\
\lt\{{p}_j(e),{h}(e')\rt\} &=&i\frac{\kappa}{a_e^2} \delta_{e,e'} \frac{\t_j}{2} {h}(e')\nonumber\\
\lt\{{p}_j(e),{p}_k(e')\rt\}&=&-i\frac{\kappa}{a_e^2} \delta_{e,e'} f_{jkl} {p}_l(e')\nonumber
\ee

The complexifier technique in LQG is motivated by the coherent state construction for compact Lie groups \cite{hall}. For separable Hilbert space, the coherent state transformation (which will be seen shortly) defined on a compact Lie group in \cite{hall} is a unitary transformation from $L^2(G)$ to the BargmannÐSegalÐFock representation $L^2(G^{\mathbb{C}})\bigcap Hol(G^{\mathbb{C}})$, where $Hol(G^{\mathbb{C}})$ is the space of holomorphic functions on the complexified Lie group $G^{\mathbb{C}}$. We first proceed the construction on one copy of $\ch_e$, which is isomorphic to the separable Hilbert space $L^2(G)$ with the Haar measure on the compact Lie group. We consider the phase space $\cm_e=T^*G$ and define a, so called, complexifier as a positive function on the the phase space $\cm_e$:
\be
C_e:=\frac{a_e^2}{2\kappa}\delta^{ij}{p}_i(e){p}_j(e)
\ee
This complexifier complexifies the holonomies $h(e)$ by the following canonical transformation:
\be
g(e)=\sum_{n=1}^\infty\frac{(-i)^n}{n!}\lt\{C_e,h(e)\rt\}_{(n)}=e^{-ip_j(e)\t_j/2}h(e)\nonumber
\ee
$g(e)$ belongs to the complexified group $G^\mathbb{C}\simeq T^*G$, which provides a complex polarization of the phase space $\cm_e$. And $\lt(g(e),\overline{g(e)}\rt)$ are complex coordinates on the phase space $\cm_e$.

Then we come to the quantization, we define a quantum complexifier $\hat{C}_e$ by
\be
\hat{C}_e:=\frac{a_e^2}{2\kappa}\delta^{ij}\hat{p}_i(e)\hat{p}_j(e)=-\frac{\hbar}{2}\ t_e\Delta_e
\ee
where $\hat{p}_j(e)=it_eX^e_j/2$ and $X^e_j$ are the right-invariant vector fields on $G$, $\Delta_e=\delta^{ij}X^e_iX^e_j/4$ is the Laplacian on $G$, and $t_e=\ell_p^2/a_e^2$ is a dimension-free classicality parameter. The coherent state on edge $e$ is defined by a heat kernel evolution of the delta function followed by an analytic contunuation:
\be
\psi^{t_e}_{g(e)}\lt(h(e)\rt)&:=&\lt[e^{-\frac{\hat{C}_e}{\hbar}}\delta_{h'(e)}\lt(h(e)\rt)\rt]_{h'(e)\to g(e)}=\lt[e^{t_e\Delta_e/2}\delta_{h'(e)}\lt(h(e)\rt)\rt]_{h'(e)\to g(e)}\nonumber\\
&=&\sum_{j_e}(2j_e+1)\ e^{-t_ej_e(j_e+1)/2}\chi_{j_e}\lt(g(e)h(e)^{-1}\rt)\label{coherent}
\ee
where $g(e)$ is the complexified holonomy $g(e)=e^{-ip_j(e)\t_j/2}h(e)\in G^{\bbc}$. And the above heat kernel evolution followed by analytic continuation $h'(e)\to g(e)$ is called the coherent state transformation from $L^2$ representation to the BargmannÐSegalÐFock representation. Note that the above analytic continuation is trivially carried out, because the representation matrix elements of a compact Lie group $G$ is the polynomials of $h(e)$ in its fundamental representation. The label $g(e)$ in the coherent state $\psi^{t_e}_{g(e)}$ denotes the point of the complexified group $G^\bbc\simeq\cm_e$, where the coherent state is supposed to approximate. Moreover, it is easy to check that the coherent state $\psi^{t_e}_{g(e)}$ is an eigenstate of the annihilation operator $\hat{g}(e)$ with eigenvalue $g(e)$
\be
\hat{g}(e):=\sum_{n=1}^\infty\frac{(-i)^n}{n!(i\hbar)^n}\lt[\hat{C}_e,\hat{h}(e)\rt]_{(n)}=e^{-\frac{\hat{C}_e}{\hbar}}\hat{h}(e)e^{\frac{\hat{C}_e}{\hbar}},\ \ \ \ \ \ \hat{g}(e)\psi^{t_e}_{g(e)}=g(e)\psi^{t_e}_{g(e)}.
\ee
It is important for us that the normalized coherent states $\tilde{\psi}^t_g$ form an over-complete basis in $\ch_e$, i.e.
\be
\int_{G^\mathbb{C}}\rmd g(e)\ |\tilde{\psi}^{t_e}_{g(e)}\rangle\langle\tilde{\psi}^{t_e}_{g(e)}|=1,\ \ \ \ \ \ \ \rmd g=\frac{1}{t^3}\rmd\mu_H(h)\rmd^3p+o(t^\infty)\label{RI}
\ee
The coherent state on the full kinematical Hilbert space $\ch_{Kin}$ is defined by a infinite tensor product
\be
\tilde{\psi}_g^t=\prod_{e\in E(\g)}\tilde{\psi}^{t_e}_{g(e)}\label{coherent1}
\ee
The label $g$ means that the coherent state $\tilde{\psi}_g^t$ is peaked at the phase space point $g=g\lt(\{h(e)\}_{e\in E(\g)}, \{p_j(e)\}_{e\in E(\g)}\rt)$ in the graph-dependent phase space $\cm_{X(\g)}=[T^*G]^{|E(\g)|}$ \cite{GCS}. For the operators $\mathrm{Pol}\lt(\{\hat{h}(e)\}_{e\in E(\g)}, \{\hat{p}_j(e)\}_{e\in E(\g)}\rt)$ which is the polynomials of the holonomies and fluxes, the coherent state $\tilde{\psi}_g^t$ demonstrates the following semiclassical properties \cite{GCS}:
\be
\lag\tilde{\psi}_g^t\lt|\ \mathrm{Pol}\lt(\{\hat{h}(e)\}_{e\in E(\g)}, \{\hat{p}_j(e)\}_{e\in E(\g)}\rt)\ \rt|\tilde{\psi}_g^t\rag=\mathrm{Pol}\lt(\{{h}(e)\}_{e\in E(\g)}, \{{p}_j(e)\}_{e\in E(\g)}\rt)+o(t)
\ee
More importantly, the semiclassical behavior of non-polynomial operators e.g. the volume operator $\hat{V}(R)$ and the master constraint operator $\M$ are also analyzed by the expectation value of the coherent state $\tilde{\psi}_g^t$, by employing the semiclassical perturbation theory \cite{AQG,ce}. The results for the volume operator and the master constraint operator on a cubic graph are\footnote{$o(t)$ is the quantum fluctuation which vanishes when $\hbar\to0$}:
\be
\lag\tilde{\psi}_g^t\lt|\ \hat{V}(R)\ \rt|\tilde{\psi}_g^t\rag&=&V(R)[g]+o(t)\nonumber\\
\lag\tilde{\psi}_g^t\lt|\ \M\ \rt|\tilde{\psi}_g^t\rag&=&\mathbf{M}[g]+o(t)\label{classM}
\ee
which demonstrate the correct semiclassical limits of both operators on a cubic graph \cite{AQG,ce}. This result shows that the coherent states on a cubic graph $\tilde{\psi}_g^t$ is qualified to be the semiclassical states for AQG.

\section{The path-integral of the master constraint}\label{main}

%In this section, we restrict the analysis in a single strong equivalence class sector of the infinite tensor product Hilbert space, i.e. the strong equivalence class sector $\ch_{AL}$ which contain the vacuum state $\o=1$. A dense domain in $\ch_{AL}$ is spanned by the states each of which only has excitations on a finite number of edges. Because of the similarity between $\ch_{AL}$ and the kinematical Hilbert space of LQG, the discussion in this section can be trivial generalized to the case of LQG. Moreover, because the master constraint operator we consider here is non-graph changing, effectively the graph $\g$ we use in this section is only a finite subgraph of the infinite algebraic graph where we define AQG.

Now we consider the practical computation for the group averaging inner product for defined in Definition \ref{GA} and derive a path-integral formula from the master constraint operator $\M$ on $\ch_{Kin}$.
\begin{eqnarray}
\lag\eta(f)\Big|\eta(f')\rag_\O:=\lim_{\eps\to0}\frac{\int_{-\infty}^{\infty}\frac{\rmd\t}{2\pi}\lag f\lt|\exp\Big[i\t \big(\M-\eps\big)\Big]\rt|f'\rag_{Kin}}{\int_{-\infty}^{\infty}\frac{\rmd\t}{2\pi}\lag \O\lt|\exp\Big[i\t \big(\M-\eps\big)\Big]\rt|\O\rag_{Kin}}\label{inner}
\end{eqnarray}
where $f$,$f'$ are kinematical states in a dense domain of $\ch_{Kin}$ and $\M$ is the master constraint operator defined in Eq.(\ref{M}).

We multiply both the numerator and denominator of Eq.(\ref{inner}) by the (infinite) constant
\be
\int[D\l(u)]\ \delta\lt(\int_{-T}^T\rmd u\ \l(u)-\t\rt)
\ee
where $[D\l]$ is a measure on the infinite dimensional space of continuous functions $\{\l(u)\}_{u\in[-T,T]}$. Note that here the integral $\int_{-T}^T\rmd u\ \l(u)$ can be written as a Riemann sum $\lim_{N\to\infty}\frac{T}{N}\sum_{n=-N}^{N-1}\l_n$ with $\l_n=\l(nT/N)$ when $N\to\infty$, provided $\l(t)$ is a continuous function.

Insert this into Eq.(\ref{inner}) and perform the integral $\int\rmd\t$
\be
\lag\eta(f)\Big|\eta(f')\rag_\O&=&\lim_{\eps\to0}\frac{\int_{-\infty}^{\infty}\frac{\rmd\t}{2\pi}\lag f\lt|\exp\Big[i\t \big(\M-\eps\big)\Big]\rt|f'\rag_{Kin}\lt[\int[D\l(u)]\ \delta\lt(\int_{-T}^T\rmd u\ \l(t)-\t\rt)\rt]}{\int_{-\infty}^{\infty}\frac{\rmd\t}{2\pi}\lag \O\lt|\exp\Big[i\t \big(\M-\eps\big)\Big]\rt|\O\rag_{Kin}\lt[\int[D\l(u)]\ \delta\lt(\int_{-T}^T\rmd u\ \l(u)-\t\rt)\rt]}\nonumber\\
&=&\lim_{\eps\to0}\frac{\int[D\l(u)]\lag f\lt|\exp\Big[\int_{-T}^T\rmd t\ \l(u)\ \big(\M-\eps\big)\Big]\rt|f'\rag_{Kin}}{\int[D\l(u)]\lag \O\lt|\exp\Big[\int_{-T}^T\rmd u\ \l(u)\ \big(\M-\eps\big)\Big]\rt|\O\rag_{Kin}}\label{inner1}
\ee

Therefore the central computation is for the matrix element
\be
\lag f\lt|\exp\Big[\int_{-T}^T\rmd u\ \l(u)\ \big(\M-\eps\big)\Big]\rt|f'\rag_{Kin}
\ee
by given a continuous function $\l(u)$ on the interval $[-T,T]$. By write the integral $\int_{-T}^T\rmd u\ \l(u)$ as a Riemann sum:
\be
\lag f\lt|\exp\Big[\int_{-T}^T\rmd u\ \l(u)\ \big(\M-\eps\big)\Big]\rt|f'\rag_{Kin}&=&\lim_{N\to\infty}\lag f\lt|\exp\Big[\sum_{n=-N}^{N-1}\frac{i T}{N}\l_n \big(\M-\eps\big)\Big]\rt|f'\rag_{Kin}\nonumber\\
&=&\lim_{N\to\infty}\lag f\lt|\prod_{n=-N}^{N-1}\exp\Big[\frac{i T}{N}\l_n \big(\M-\eps\big)\Big]\rt|f'\rag_{Kin}
\ee
Then we insert in the resolution of identity with coherent states Eq.(\ref{RI}) on $\ch_{Kin}$,
\be
\prod_{e\in E(\g)}\int_{G^\mathbb{C}}\rmd g(e)\ |\tilde{\psi}^{t_e}_{g(e)}\rangle\langle\tilde{\psi}^{t_e}_{g(e)}|=1
\ee
we obtain
\be
\lag f\lt|\exp\Big[\int_{-T}^T\rmd t\ \l(t)\ \big(\M-\eps\big)\Big]\rt|f'\rag_{Kin}=\lim_{N\to\infty}\int\prod_{n=-N}^{N}\mathrm{d}g_{n}\prod_{n=-N}^{N-1}\lag \tilde{\psi}^t_{g_{n+1}}\lt|\exp\Big[\frac{i T}{N}\l_n \big(\M-\eps\big)\Big]\rt|\tilde{\psi}^t_{g_{n}}\rag_{Kin}\lag f\ \Big|\  \tilde{\psi}^t_{g_N}\rag_{Kin}\lag \tilde{\psi}^t_{g_{-N}}\ \Big|\ f'\rag_{Kin}\label{applykato}
\ee
Here $\rmd g_n=\prod_{e\in E(\g)}\rmd\mu_H(h_n)\rmd^3p_n/{t^3}$ is the Liouville measure on the graph phase space $T^*G^{|E(\g)|}$.

When we take the limit $N\to \infty$, we could approximate the matrix elements (as we usually do in the standard textbook-derivation of path integral formula)
\be
\lag \tilde{\psi}^t_{g_{n+1}}\lt|\exp\Big[\frac{i T}{N}\l_n \big(\M-\eps\big)\Big]\rt|\tilde{\psi}^t_{g_{n}}\rag_{Kin}\simeq\lag \tilde{\psi}^t_{g_{n+1}}\lt|1+\frac{i T}{N}\l_n \big(\M-\eps\big)\rt|\tilde{\psi}^t_{g_{n}}\rag_{Kin}
\ee
Here $\l(t)$ is continuous function thus is bounded, so $\frac{T}{N}\l_n$ is very small. Of course such an approximation still relies on non-trivial assumptions on the properties of the Master constraint operator (e.g. some assumption described in \cite{klauder}).

First let's compute the single-step amplitude $\langle \tilde{\psi}^t_{g_{i}}\big|1+\frac{i T}{N}\l_n\big(\mathbf{M}-\epsilon\big)\big|\tilde{\psi}^t_{g_{i-1}}\rangle_{Kin}$. Here the overlap function $\langle \tilde{\psi}^t_{g_{i}}|\tilde{\psi}^t_{g_{i-1}}\rangle_{Kin}$ is sharply peaked at $g_i=g_{i-1}$ in a Gaussian fashion (with width $\sqrt{t}$), Thus in the semiclassical limit $t\rightarrow0$ we have
\begin{eqnarray}
&&\lim_{t\rightarrow0}\lag \tilde{\psi}^t_{g_{i}}\lt|1+\frac{i T}{N}\l_n\big(\mathbf{M}-\epsilon\big)\rt|\tilde{\psi}^t_{g_{i-1}}\rag_{Kin}\ =\ \lim_{t\rightarrow0}\Bigg[1+\frac{i T}{N}\l_n\frac{\langle \tilde{\psi}^t_{g_{i}}|\mathbf{M}-\epsilon|\tilde{\psi}^t_{g_{i-1}}\rangle_{Kin}}{\langle \tilde{\psi}^t_{g_{i}}|\tilde{\psi}^t_{g_{i-1}}\rangle_{Kin}}\Bigg]\langle \tilde{\psi}^t_{g_{i}}|\tilde{\psi}^t_{g_{i-1}}\rangle_{Kin}\nonumber\\
&=&\lim_{t\rightarrow0}\Bigg[1+\frac{i T}{N}\l_n\frac{\langle \tilde{\psi}^t_{g_{i}}|\mathbf{M}-\epsilon|\tilde{\psi}^t_{g_{i}}\rangle_{Kin}}{\langle \tilde{\psi}^t_{g_{i}}|\tilde{\psi}^t_{g_{i}}\rangle_{Kin}}\Bigg]\langle \tilde{\psi}^t_{g_{i}}|\tilde{\psi}^t_{g_{i-1}}\rangle_{Kin}\ =\ \lim_{t\rightarrow0}\Bigg\{1+\frac{i T}{N}\l_n\Big(\mathbf{M}[g_i]-\epsilon\Big)\Bigg\}\langle \tilde{\psi}^t_{g_{i}}|\tilde{\psi}^t_{g_{i-1}}\rangle_{Kin}\nonumber
\end{eqnarray}
In the second step, we use the fact that as $t\to0$, the overlap function $\langle \tilde{\psi}^t_{g_{i}}|\tilde{\psi}^t_{g_{i-1}}\rangle_{Kin}$ is proportional to a delta function $\delta_{g_i}(g_{i-1})$. In the last step we use the fact that the master constraint operator has correct semiclassical limit Eq.(\ref{classM}). Therefore, we have seen that
\begin{eqnarray}
\lag\tilde{\psi}^t_{g_{i}}\lt|1+\frac{i T}{N}\l_n\big(\mathbf{M}-\epsilon\big)\rt|\tilde{\psi}^t_{g_{i-1}}\rag_{Kin}\ =\ \lt[1+\frac{i T}{N}\l_n\lt(\mathbf{M}[g_i]-\epsilon+tF^t(g_i,g_{i-1})\rt)\rt]\langle \tilde{\psi}^t_{g_{i}}|\tilde{\psi}^t_{g_{i-1}}\rangle_{Kin}\nonumber
\end{eqnarray}
where $F^t(g_i,g_{i-1})$ is the fluctuation of the master constraint operator $\M$ with respect to the coherent states $\tilde{\psi}^t_{g_{i}}$ and $\tilde{\psi}^t_{g_{i-1}}$. Therefore as we formally take that limit $N\to\infty$, we have approximately
\begin{eqnarray}
\lag \tilde{\psi}^t_{g_{n+1}}\lt|\exp\Big[\frac{i T}{N}\l_n \big(\M-\eps\big)\Big]\rt|\tilde{\psi}^t_{g_{n}}\rag_{Kin}\simeq \exp\lt[\frac{i T}{N}\l_n \lt(\mathbf{M}[g_i]-\epsilon+tF^t(g_i,g_{i-1})\rt)\rt]\langle \tilde{\psi}^t_{g_{i}}|\tilde{\psi}^t_{g_{i-1}}\rangle_{Kin}.
\end{eqnarray}
Then let's evaluate the overlap function $\langle \tilde{\psi}^t_{g_{i}}|\tilde{\psi}^t_{g_{i-1}}\rangle_{Kin}=\prod_{e\in E(\g)}\langle {\psi}^t_{g_{i}(e)}|{\psi}^t_{g_{i-1}(e)}\rangle_{Kin}/||{\psi}^t_{g_{i}(e)}||\ ||{\psi}^t_{g_{i-1}(e)}||$ where
\be
{\psi}^t_{g(e)}\lt(h(e)\rt)=\sum_{2j_e=0}^\infty(2j_e+1)e^{-tj_e(j_e+1)/2}\chi_{j_e}\Big(g(e)h^{-1}(e)\Big)
\ee
is the complexifier coherent state on the edge $e$. If we set $n=2j+1\in\mathbb{N}_0$, on one edge
\begin{eqnarray}
\langle {\psi}^t_{g_{i}}|{\psi}^t_{g_{i-1}}\rangle_{Kin}=\frac{e^{{t}/{4}}}{2\sinh(z_{i,i-1})}\sum_{n\in\mathbb{Z}}n\ e^{tn^2/4}e^{nz_{i,i-1}}
\end{eqnarray}
where $\cosh(z_{ij})=\mathrm{tr}(g_i^\dagger g_j)/2$. For the norms of $\psi^t_{g_j}$, the same formula holds just with $z_{ij}$ replaced by $z_i$ such that $\cosh(z_i)=\mathrm{tr}(g_i^\dagger g_i)/2$. Therefore,
\begin{eqnarray}
\langle \tilde{\psi}^t_{g_{i}}|\tilde{\psi}^t_{g_{i-1}}\rangle_{Kin}&=&\prod_{e\in E(\gamma)}\frac{\frac{e^{{t}/{4}}}{2\sinh(z_{i,i-1})}\sum_{n\in\mathbb{Z}}n\ e^{tn^2/4}e^{nz_{i,i-1}}}{\sqrt{\frac{e^{{t}/{4}}}{2\sinh(z_{i})}\sum_{n\in\mathbb{Z}}n\ e^{tn^2/4}e^{nz_{i}}}\sqrt{\frac{e^{{t}/{4}}}{2\sinh(z_{i-1})}\sum_{n\in\mathbb{Z}}n\ e^{tn^2/4}e^{nz_{i-1}}}}\nonumber\\
&\simeq&\prod_{e\in E(\gamma)}\frac{\sqrt{|\sinh(z_{i})\sinh(z_{i-1})|}}{\sinh(z_{i,i-1})}\frac{z_{i,i-1}e^{z_{i,i-1}^2/t}}{\sqrt{|z_{i}e^{z_{i}^2/t}|}\sqrt{|z_{i-1}e^{z_{i-1}^2/t}}|}\nonumber\\
&=&\prod_{e\in E(\gamma)}\frac{z_{i,i-1}\sqrt{|\sinh(z_{i})\sinh(z_{i-1})|}}{\sqrt{|z_iz_{i-1}|}\sinh(z_{i,i-1})}e^{[z_{i,i-1}^2-\frac{1}{2}z_i^2-\frac{1}{2}z_{i-1}^2]/t}\label{overlap}
\end{eqnarray}
where we have neglect a term of order $o(t^\infty)$ in the second step, and we have used the Poisson resummation formula
\be
\sum_{n\in\mathbb{Z}}f(ns)=\frac{1}{s}\sum_{n\in\mathbb{Z}}\int_{\mathbb{R}}\rmd x\ e^{2\pi inx/s}f(x)
\ee
with $s=\sqrt{t_e}$.

On the other hand, we have the relation:
\begin{eqnarray}
\cosh(z_i)&=&\mathrm{tr}(g_ig_i^\dagger)/2\ =\ \mathrm{tr}(h_i^\dagger e^{-ip_i\cdot\tau/2}e^{-ip_i\cdot\tau/2}h_i)/2\nonumber\\
&=&\mathrm{tr}(e^{-ip_i\cdot\tau})/2\ =\ \cosh(p_i)
\end{eqnarray}
which gives $|z_i|=p_i\equiv\sqrt{p_i\cdot p_i}$.

Then we can apply the above result for the overlap function to compute the matrix element (as $N\to\infty$)
\begin{eqnarray}
&&\lag f\lt|\exp\Big[\int_{-T}^T\rmd u\ \l(u)\ \big(\M-\eps\big)\Big]\rt|f'\rag_{Kin}\nonumber\\
&=&\lim_{N\to\infty}\int\prod_{e\in E(\gamma)}\prod_{i=-N}^{N}\frac{\mathrm{d}h_i\mathrm{d}^3p_i}{t^3}\frac{\sinh(p_i)}{p_i}e^{-p_i^2/t}\nonumber\\
&&\prod_{k=-N+1}^{N}\frac{z_{k,k-1}}{\sinh(z_{k,k-1})}e^{z^2_{k,k-1}/t} \exp\lt[\frac{i T}{N}\l_k\lt(\mathbf{M}[g_k]-\epsilon+tF^t(g_k,g_{k-1})\rt)\rt]\ \overline{f(g_N)}f'(g_{-N})\label{i0}
\end{eqnarray}
where $f(g):=\langle \tilde{\psi}^t_{g}\ |\ f\rangle_{Kin}$.

%Note that since we are working in the strong equivalent sector $\ch_{AL}$, the graph $\g$ is essentially a finite algebraic cubic graph with a finite number of edges. Similar to the spin-foam models, the path-integral formula Eq.(\ref{i0}) is featured by a discrete foam structure where the elementary cell is a hypercube. However, the structures in the master constraint $\bf{M}$ and the fluctuation $F^t$ displays many non-trivial interactions between different hypercubes. In addition, $\overline{f(g_N)}f'(g_0)$ is the boundary state of the model but expressed in the coherent state representation.

Furthermore, we define the Lie algebra variables $\theta_k$ such that $h_k=e^{\theta_k\cdot\tau/2}$, $z_{k,k-1}$ can be computed by
\begin{eqnarray}
\cosh(z_{k,k-1})&=&\frac{1}{2}\mathrm{tr}(h_{k-1}h_k^\dagger e^{-i(p_k+p_{k-1})\cdot\tau/2})\nonumber\\
&=&\frac{1}{2}\mathrm{tr}(e^{i[-(p_k+p_{k-1})+i(\theta_k-\theta_{k-1})]\cdot\tau/2})\nonumber\\
&=&\cosh\Bigg(\sqrt{\Big[-\frac{(p_k+p_{k-1})}{2}+i\frac{(\theta_k-\theta_{k-1})}{2}\Big]^2}\ \Bigg)
\end{eqnarray}
Therefore,
\begin{eqnarray}
z_{k,k-1}^2-\frac{1}{2}p_k^2-\frac{1}{2}p_{k-1}^2=-\frac{1}{4}\Big[(p_k-p_{k-1})^2+(\theta_k-\theta_{k-1})^2+2i(p_k+p_{k-1})\cdot(\theta_k-\theta_{k-1})\Big]
\end{eqnarray}
Then we insert the result back into Eq.(\ref{i0}) and obtain
\be
&&\lag f\lt|\exp\Big[\int_{-T}^T\rmd u\ \l(u)\ \big(\M-\eps\big)\Big]\rt|f'\rag_{Kin}\nonumber\\
&=&\lim_{N\to\infty}\int\prod_{e\in E(\gamma)}\prod_{i=-N}^{N}\frac{\mathrm{d}h_i\mathrm{d}^3p_i}{t^3}\frac{\sinh(p_i)}{p_i}e^{-(p^2_{-N}+p^2_N)/2t}\prod_{k=-N+1}^N\frac{z_{k,k-1}}{\sinh(z_{k,k-1})}\ \overline{f(g_N)}f'(g_{-N})\nonumber\\
&&\times\exp\lt\{-i\frac{(p_k+p_{k-1})}{2t}\cdot(\theta_{k}-\theta_{k-1})-\frac{1}{4t}\Big[(p_k-p_{k-1})^2+(\theta_k-\theta_{k-1})^2\Big]+
\frac{i T}{N}\l_k\lt[\mathbf{M}[g_k]-\epsilon+tF^t(g_k,g_{k-1})\rt]\rt\}\label{i1}
\ee
In order to contact with the path-integral formula with a classical action on the exponential, we make the following approximations for Eq.(\ref{i1})\footnote{These approximations essentially remove all the $t^{n\geqslant0}$-order contributions on the exponential, and keep only the $t^{-1}$-order contributions.}:

\begin{itemize}
\item We assume the fluctuation $F^t$ of the master constraint operator $\M$ with respect to the coherent states $\psi^t_g$ is small and negligible. It is a non-trivial assumption for the property of the master constraint operator $\M$, i.e. we should design a certain operator ordering in the definition of the self-adjoint master constraint operator $\M$, such that the fluctuation $F^t$ is small and negligible. For simple system like the ordinary free quantum field theory, such an operator ordering is nothing but the normal ordering of the creation and annihilation operator, which results in $F^t=0$.

\item We only count the paths such that the second order terms $(\Delta p)^2$ and $(\Delta\theta)^2$ on the exponential is negligible. Note that the overlap function $\langle \tilde{\psi}^t_{g_{i}}|\tilde{\psi}^t_{g_{i-1}}\rangle_{Kin}$ is sharply peaked at the point $g_i=g_{i-1}$ with width $\sqrt{t}$. Thus $(\Delta p)^2,(\Delta\theta)^2\sim t$ and have to be neglected if we want to have a classical action on the exponential. It is consistent if we want to have a classical kinetic term containing the time-derivative of $\theta$ after the continuum limit.
\end{itemize}

With the above approximations. Eq.(\ref{i1}) is simplified to be
\begin{eqnarray}
&&\lag f\lt|\exp\Big[\int_{-T}^T\rmd u\ \l(u)\ \big(\M-\eps\big)\Big]\rt|f'\rag_{Kin}\nonumber\\
&=&\lim_{N\to\infty}\int\prod_{e\in E(\gamma)}\prod_{i=-N}^{N}\frac{\mathrm{d}h_i\mathrm{d}^3p_i}{t^3}\frac{\sinh(p_i)}{p_i}e^{-(p^2_{-N}+p^2_N)/2t}
\prod_{k=-N+1}^N\frac{z_{k,k-1}}{\sinh(z_{k,k-1})}\ \overline{f(g_N)}f'(g_{-N})\nonumber\\
&&\prod_{k=-N+1}^N\exp\lt\{-i\frac{(p_k+p_{k-1})}{2t}\cdot(\theta_{k}-\theta_{k-1})+
\frac{i T}{N}\l_k\lt[\mathbf{M}[g_k]-\epsilon\rt]\rt\}\nonumber\\
&=&\int\lt[D\theta^j_e(x^0) Dp_j^e(x^0)\rt]\ \exp\lt\{-\frac{i}{t}\int_{-T}^T\rmd x^0\sum_{e\in E(\g)}p_j^e\partial_{x^0}{\theta}^j_{e}+
i\int_{-T}^T\rmd x^0\ \l(x^0)\lt[\mathbf{M}[p^e_j(x^0),\theta_e^j(x^0)]-\epsilon\rt]\rt\}\nonumber\\ &&\overline{f\lt[p^e_j(T),\theta_e^j(T)\rt]}f'\lt[p^e_j(-T),\theta_e^j(-T)\rt]\label{i2}
\end{eqnarray}
where we have relabel the coordinate $u=x^0$, and the functional measure is formally defined by
\be
\lt[D\theta^j_e(x^0) Dp_j^e(x^0)\rt]:=\lim_{N\to\infty}\prod_{e\in E(\gamma)}\prod_{i=-N}^{N}\frac{\mathrm{d}h_i\mathrm{d}^3p_i}{t^3}\frac{\sinh(p_i)}{p_i}e^{-(p^2_{-N}+p^2_N)/2t}
\prod_{k=-N+1}^N\frac{z_{k,k-1}}{\sinh(z_{k,k-1})}.
\ee

Then we insert Eq.(\ref{i2}) back to Eq.(\ref{inner1}) and perform the functional integral of $D\l$ (we consider the numerator)
\begin{eqnarray}
&&\lim_{\eps\to0}\int[D\l(u)]\lag f\lt|\exp\Big[\int_{-T}^T\rmd t\ \l(u)\ \big(\M-\eps\big)\Big]\rt|f'\rag_{Kin}\nonumber\\
&=&\lim_{\epsilon\to0}\int[D\l(x^0)]\int\lt[D\theta^j_e(x^0) Dp_j^e(x^0)\rt]\ \exp\lt\{-\frac{i}{t}\int_{-T}^T\rmd x^0\sum_{e\in E(\g)}p_j^e\partial_{x^0}{\theta}^j_{e}+
i\int_{-T}^T\rmd x^0\ \l(x^0)\lt[\mathbf{M}[p^e_j(x^0),\theta_e^j(x^0)]-\epsilon\rt]\rt\}\nonumber\\ &&\overline{f\lt[p^e_j(T),\theta_e^j(T)\rt]}f'\lt[p^e_j(-T),\theta_e^j(-T)\rt]\nonumber\\
&=&\lim_{\epsilon\to0}\int\lt[D\theta^j_e(x^0) Dp_j^e(x^0)\rt]\ \exp\lt\{-\frac{i}{t}\int_{-T}^T\rmd x^0\sum_{e\in E(\g)}p_j^e\partial_{x^0}{\theta}^j_{e}\ \rt\}\prod_{x^0\in[-T,T]}\delta\lt(\mathbf{M}[p^e_j(x^0),\theta_e^j(x^0)]-\epsilon\rt)\nonumber\\ &&\overline{f\lt[p^e_j(T),\theta_e^j(T)\rt]}f'\lt[p^e_j(-T),\theta_e^j(-T)\rt]\label{masterPI}
\end{eqnarray}

Now we consider the delta function $\delta\lt(\mathbf{M}[p^e_j(x^0),\theta_e^j(x^0)]-\epsilon\rt)$ for each $x^0$. If the graph $\g$ here was a finite graph with a finite number of vertices, thus the sum in master constraint $\sum_{v\in E(\g)}$ would be a finite sum. However, even we consider an infinite graph with infinite vertices, i.e. the master constraint has a shape $\mathbf{M}=\sum_{I\in\ci}C_IC_I$ with some constraints $C_I$, where the sum in the expression is a infinite sum and $\ci$ is a infinite index set. We can truncate the sum and define a partial master constraint $\mathbf{M}_W=\sum_{I\in W}C_IC_I$ where $W\subset\ci$ is a finite set. Then it turns out that the group averaging inner product by using the partial master constraint $\mathbf{M}_W$ converges to the group averaging inner product by using full master constraint $\mathbf{M}$ as $W\to\ci$ under some certain technique assumptions (see \cite{link1} for the details). It means that for a infinite graph $\g$ we could define a delta function for the partial master constraint,
\be
\delta_W(\mathbf{M}-\eps):=\delta(\mathbf{M}_W-\eps)\ \ \ \ \text{such that}\ \ \ \ \lim_{W\to\ci}\delta_W(\mathbf{M}-\eps)=\delta(\mathbf{M}-\eps)
\ee
in the sense of distribution.

So here we only consider the following type of integral with a finite integer $N$ and any test function $f$ continuous in a neighborhood of the constraint surface:
\be
&&\lim_{\epsilon\to0}\int\mathrm{d}x_1\mathrm{d}x_2\cdots\mathrm{d}x_N\ \delta(\sum_{i=1}^Nx_ix_i-\epsilon)\ f(\vec{x})\nonumber\\
&=&\lim_{\epsilon\to0}\int\mathrm{d}x_2\cdots\mathrm{d}x_N\ \frac{1}{2\sqrt{\epsilon-\Sigma_{i=2}^Nx_ix_i}}\ \Bigg[f\Big(x_1=\sqrt{\epsilon-\Sigma_{i=2}^Nx_ix_i}\Big)+f\Big(x_1=-\sqrt{\epsilon-\Sigma_{i=2}^Nx_ix_i}\Big)\Bigg]\nonumber\\
&=&\Bigg\{\lim_{\epsilon\to0}\int\mathrm{d}x_2\cdots\mathrm{d}x_N\ \frac{1}{2\sqrt{\epsilon-\Sigma_{i=2}^Nx_ix_i}}\Bigg\}\ f\big(\vec{x}=0\big)\nonumber\\
&\equiv&\lim_{\epsilon\to0}\mathcal{N}(\epsilon)f\big(\vec{x}=0\big)\nonumber\\
&=&\lim_{\epsilon\to0}\mathcal{N}(\epsilon)\int\mathrm{d}x_1\mathrm{d}x_2\cdots\mathrm{d}x_N\prod_{i=1}^N \delta(x_i)\ f(\vec{x})\nonumber
\ee
where $\mathcal{N}(\epsilon)$ diverges in the limit $\epsilon\to0$.

So far for the integral on the space of $\vec{x}$, if we parametrize $\{x_i\}_{i=1}^N$ by the variables $\{t_j\}_{j=1}^M$ ($M>N$) and consider the following integral:
\be
\lim_{\epsilon\to0}\int\mathrm{d}t_1\cdots\mathrm{d}t_M\ \delta(\sum_{i=1}^Nx_i(\vec{t})x_i(\vec{t})-\epsilon)\ f(\vec{t}).\nonumber
\ee
Then we make the changing of variables
\be
t_k&\to& x_k(\vec{t})\ \ \ \ \ \ \ \mathrm{for}\ \ k\in\{1,\cdots,N\}\nonumber\\
t_k&\to& t_k\ \ \ \ \ \ \ \ \ \ \ \ \mathrm{for}\ \ k\in\{N+1,\cdots,M\}\nonumber
\ee
and denote the Jacobian to be $\det(\partial x/\partial t)$ and assume it is nonzero, continuous and bounded around the constraint surface. Therefore
\be
&&\lim_{\epsilon\to0}\int\mathrm{d}t_1\cdots\mathrm{d}t_M\ \delta(\sum_{i=1}^Nx_i(\vec{t})x_i(\vec{t})-\epsilon)\ f(t_1,\cdots,t_M)\nonumber\\
&=&\lim_{\epsilon\to0}\int\rmd x_1\cdots\rmd x_N\rmd t_{N+1}\cdots\rmd t_M\frac{1}{\det(\partial x/\partial t)} \ \delta(\sum_{i=1}^Nx_ix_i-\epsilon)\ \tilde{f}(x_1,\cdots,x_N,t_{N+1},\cdots,t_M)\nonumber\\
&=&\lim_{\epsilon\to0}\mathcal{N}(\epsilon)\int\rmd x_1\cdots\rmd x_N\rmd t_{N+1}\cdots\rmd t_M\frac{1}{\det(\partial x/\partial t)} \prod_{i=1}^N\delta(x_i)\ \tilde{f}(x_1,\cdots,x_N,t_{N+1},\cdots,t_M)\nonumber\\
&=&\lim_{\epsilon\to0}\mathcal{N}(\epsilon)\int\rmd t_1\cdots\rmd t_M \prod_{i=1}^N\delta(x_i)\ f(t_1,\cdots,t_M).\nonumber
\ee
Therefore we obtain the result:
\be
\lim_{\epsilon\to0}\delta(\sum_{i=1}^Nx_i(\vec{t})x_i(\vec{t})-\epsilon)=\lim_{\epsilon\to0}\mathcal{N}(\epsilon)\prod_{i=1}^N\delta(x_i)
\ee

Since the master constraint in our case has the following expression:
\be
\mathbf{M}=\sum_{v\in V(\g)}\frac{G_{j,v}}{V^{1/2}_v}\frac{G_{j,v}}{V^{1/2}_v}+\frac{D_{j,v}}{V^{1/2}_v}\frac{D_{j,v}}{V^{1/2}_v}+\frac{H_v}{V^{1/2}_v}\frac{H_v}{V^{1/2}_v}\label{M2}
\ee
where $G_j$, $D_j$ and $H$ here are the original Gauss constraint, diffeomorphism constraint and Hamiltonian constraint in their regularized forms, then we obtain the following path-integral formula
\begin{eqnarray}
&&\lim_{\eps\to0}\int[D\l(u)]\lag f\lt|\exp\Big[\int_{-T}^T\rmd t\ \l(u)\ \big(\M-\eps\big)\Big]\rt|f'\rag_{Kin}\nonumber\\
&=&\lim_{\epsilon\to0}\cn(\eps)\int\lt[D\theta^j_e(x^0) Dp_j^e(x^0)\rt]\ \exp\lt\{-\frac{i}{t}\int_{-T}^T\rmd x^0\sum_{e\in E(\g)}p_j^e\partial_{x^0}{\theta}^j_{e}\ \rt\}\overline{f\lt[p^e_j(T),\theta_e^j(T)\rt]}f'\lt[p^e_j(-T),\theta_e^j(-T)\rt]\nonumber\\
&&\prod_{x^0\in[-T,T]}\prod_{v\in V(\g)}
\delta^3\lt(\frac{G_{j,v}}{V^{1/2}_v}[p^e_j(x^0),\theta_e^j(x^0)]\rt)
\delta^3\lt(\frac{D_{j,v}}{V^{1/2}_v}[p^e_j(x^0),\theta_e^j(x^0)]\rt)
\delta\lt(\frac{H_{v}}{V^{1/2}_v}[p^e_j(x^0),\theta_e^j(x^0)]\rt)\nonumber\\
&=&\lim_{\epsilon\to0}\cn(\eps)\int\lt[D\theta^j_e(x^0) Dp_j^e(x^0) D\L_v^j(x^0) DN_v^j(x^0) D N(x^0)\rt]\ \overline{f\lt[p^e_j(T),\theta_e^j(T)\rt]}f'\lt[p^e_j(-T),\theta_e^j(-T)\rt]\nonumber\\
&&\exp\lt\{-\frac{i}{t}\int_{-T}^T\rmd x^0\lt[\sum_{e\in E(\g)}p_j^e\partial_{x^0}{\theta}^j_{e}+\sum_{v\in V(\g)}\lt(\L^j_vG_{j,v}+N^j_vD_{j,v}+N_vH_v\rt)\ \rt]\  \rt\}\label{i3}
\end{eqnarray}
where the factor $\lim_{\epsilon\to0}\cn(\eps)$ can be canceled out from both denominator and numerator of Eq.(\ref{inner1}). Here the path-integral measure is defined formally by
\be
&&\lt[D\theta^j_e(x^0) Dp_j^e(x^0) D\L_v^j(x^0) DN_v^j(x^0) D N(x^0)\rt]\nonumber\\
&:=&\lim_{N\to\infty}\lt[\prod_{i=-N}^{N}\prod_{e\in E(\gamma)}\frac{\mathrm{d}h_i\mathrm{d}^3p_i}{t^3}\frac{\sinh(p_i)}{p_i}e^{-(p^2_{-N}+p^2_N)/2t}
\prod_{k=-N+1}^N\frac{z_{k,k-1}}{\sinh(z_{k,k-1})}\rt]\lt[\prod_{n=-N+1}^N\prod_{v\in V(\gamma)}\mathrm{d}^3\Lambda_{v,n}\mathrm{d}^3N_{v,n}\mathrm{d}N_{v,n}\rt]\lt[\prod_{n=-N+1}^N\prod_{v\in V(\gamma)}V^{7/2}_v[p_n]\rt].\nonumber
\ee

In the end we write down this path-integral representation of the group averaging physical inner product
\be
&&\lag\eta(f)|\eta(f')\rag_\Omega=\frac{Z_\g(f,f')}{Z_\g(\O,\O)}\label{*}
\ee
with
\be
Z_\g(f,f')&=&\int\lt[D\theta^j_e(x^0) Dp_j^e(x^0) D\L_v^j(x^0) DN_v^j(x^0) D N(x^0)\rt]\ \overline{f\lt[p^e_j(T),\theta_e^j(T)\rt]}f'\lt[p^e_j(-T),\theta_e^j(-T)\rt]\nonumber\\
&&\exp\lt\{-\frac{i}{t}\int_{-T}^T\rmd x^0\lt[\sum_{e\in E(\g)}p_j^e\partial_{x^0}{\theta}^j_{e}+\sum_{v\in V(\g)}\lt(\L^j_vG_{j,v}+N^j_vD_{j,v}+N_vH_v\rt)\ \rt]\  \rt\}.
\ee

There are a few remarks for the path-integral formula Eq.(\ref{i3}): Motivated by the following relation between $g(e)=e^{-ip_j(e)\t_j/2}h(e)\in T^*G^{|E(\g)|}$ and the classical phase space variable $(A^j_a, E^a_j)\in\cm$ via the embedding $x^\mu:\ \g\to\Sig$
\begin{eqnarray}
h(e)&=&e^{\theta^j(e)\tau_j/2}\ =\ \mathcal{P}e^{\int_e A^j\tau^j/2}\nonumber\\
a_e^2p_j(e)&=&-\mathrm{tr}\Big(\frac{\tau_j}{2}\int_{S_e}\mathrm{d}x^a\wedge\mathrm{d}x^b\epsilon_{abc}h(\rho_e(x))E^c(x)h(\rho_e(x))^{-1}\Big)\nonumber
\end{eqnarray}
Then obviously Eq.(\ref{i3}) is in analogy with a path-integral formula on the spacetime manifold $M$ when we take a certain continuum limit (recall that $t=\ell_p^2/a^2$):
\be
&&Z_{\g\to\Sig}(f,f')\nonumber\\
&=&\int\lt[\cd A_a^j\cd E^a_j\cd\L^j\cd N^a\cd N\rt]\ \overline{f\lt[A_a^j,E^a_j\rt]_T}f'\lt[A_a^j,E^a_j\rt]_{-T}\nonumber\\
&&\times\exp\lt[-\frac{i}{\ell_p^2}\int\rmd^4x\lt(E^a_j\partial_{x^0}A^j_a-\L^jG_j-N^aH_a-NH\rt)\rt]\label{i4}
\ee
It is obvious that the integral on the exponential is nothing but the canonical action of GR in terms of su(2) connection and electric field variables. And the path integral measure is the continuum version of
\be
&&\lt[D\theta^j_e(x^0) Dp_j^e(x^0) D\L_v^j(x^0) DN_v^j(x^0) D N(x^0)\rt]\nonumber\\
&:=&\lim_{N\to\infty}\lt[\prod_{i=-N}^{N}\prod_{e\in E(\gamma)}\frac{\mathrm{d}h_i\mathrm{d}^3p_i}{t^3}\frac{\sinh(p_i)}{p_i}e^{-(p^2_{-N}+p^2_N)/2t}
\prod_{k=-N+1}^N\frac{z_{k,k-1}}{\sinh(z_{k,k-1})}\rt]\lt[\prod_{n=-N+1}^N\prod_{v\in V(\gamma)}\mathrm{d}^3\Lambda_{v,n}\mathrm{d}^3N_{v,n}\mathrm{d}N_{v,n}\rt]\lt[\prod_{n=-N+1}^N\prod_{v\in V(\gamma)}V^{7/2}_v[p_n]\rt].\nonumber
\ee
Thus the naive Ansatz of the path-integral formula Eq.(\ref{naive}) is reproduced by Eq.(\ref{i4}) up to the measure factor. The spatial volume local measure factor $V_s^{7/2}$ appeared in the path-integral measure is because the master constraint is of the type $\mathbf{M}=\sum_{I}K_{IJ}C_IC_J$ (recall Eq.(\ref{M2})), with a phase space dependent $K_{IJ}$. However for example, if we define the master constraint by
\be
\mathbf{M}=\sum_{v\in V(\g)}{G_{j,v}}{G_{j,v}}+{D_{j,v}}{D_{j,v}}+{H_v}{H_v}
\ee
instead of Eq.(\ref{M2}), the spatial volume factor $V_s^{7/2}$ disappears in the path-integral formula. However, the definition Eq.(\ref{M2}) is preferred because it has diffeomorphism invariant continuum limit.

\section{Conclusion and discussion}\label{conc}

In the previous analysis, we employ the group averaging technique for the AQG master constraint $\M$ on a cubic algebraic graph $\g$, and derive the expected path-integral formula for the group averaging physical inner product Eq.(\ref{*}), which is a certain analog of the Feynman-Kac formula in quantum mechanics \cite{reedsimon}. The derivation uses the semiclassical tools developed in \cite{AQG,GCS}, thus the final path-integral formula is in the form of a coherent state path-integral. Although the resulting path-integral formula relies on non-trivial assumptions and approximations, our analysis suggests the existence of a path-integral formulation directly relating the master constraint quantization for GR.

Another implication of the previous analysis is that, we suggest a more direct path linking canonical LQG with a path integral quantization of GR, comparing with the approaches we suggested in \cite{EHT}. The procedure in \cite{EHT} is the following: We start with the canonical theory of GR (the Holst and the Plebanski-Holst formulation) on the continuum, and perform the formal quantization in its reduced phase space. Then we derive from the reduced phase space quantization a formal path-integral of GR on the continuum, which is formally proved to represent the physical inner product of GR.

However the derivation in the present paper start from Dirac quantization in terms of a concrete operator formalism. We start with the quantization of GR via AQG on an infinite algebraic cubic graph $\g$\footnote{We can also start from LQG with a chosen graph $\g$, the derivation for the path-integral formula is the same as we did in Section \ref{main}.} and its self-adjoint master constraint operator $\M$. The physical inner product is formulated in terms of the group averaging inner product of the self-adjoint master constraint $\M$. The skeletonization procedure for the group averaging of $\M$ arrives at a path integral representation Eq.(\ref{*}) of the group averaging physical inner product from the operator formalism. Interestingly the skeletonization procedure in the derivation shows some certain similar structures between the group averaging physical inner product and the spin-foam quantization (on a hypercubic spacetime lattice). Thus it could be a future research about derive a spin-foam vertex amplitude from e.g. Eq.(\ref{i0}).

The resulting path-integral formula in the present analysis is in a canonical form Eq.(\ref{i3}), i.e. the action on the exponential is the canonical action of GR and the path-integral measure $\mathrm{d}h_i\mathrm{d}^3p_i$ is the Liouville measure on the phase space $T^*G^{|E(\g)|}$. It is interesting to consider deriving from this canonical path-integral to a covariant path-integral with original covariant field variables, where we may have to make a certain unfolding of the path-integral measure \cite{EHT} and use the Henneaux-Slavnov trick \cite{hs}. From this procedure it would be interesting to see if the SO(4) or SL(2,$\bbc$) holonomies of the spacetime covariant connection fields can appear in the final formulation. All current spin-foam models \cite{spinfoam2} start from a discretized path-integral formula in terms of the SO(4) or SL(2,$\bbc$) holonomies and the Haar measure of SO(4) or SL(2,$\bbc$), but the result here Eq.(\ref{i3}) only display a SU(2) Haar measure $\rmd h(e)$ product with the Lebesgue measure $\rmd \L_v$ standing for the spacetime connection field.

\section*{Acknowledgments}

M.H. is grateful for the advises from Thomas Thiemann and the fruitful discussion with Kristina Giesel about semiclassical analysis, and thanks the referees for their suggestions. M.H. also would like to gratefully acknowledge the support by International Max Planck Research School and the partial support by NSFC Nos. 10675019 and 10975017.

\end{document}